\documentclass{ctr}

\usepackage{ctrfont}
\usepackage{natbib}

\usepackage{url}
\usepackage{amsmath}
\usepackage{amssymb}

\usepackage{graphicx}
\usepackage{placeins}

\title{RANS predictions for high-speed flows using enveloping models}
\shorttitle{RANS predictions using enveloping models}
\author{A. A. Mishra \and G. Iaccarino}
\shortauthor{A. A. Mishra \& G. Iaccarino}

\begin{document}


\maketitle

\section{Motivation and objectives} 

In spite of over a century of research, no analytic solutions for the equations governing turbulent flows are available. Furthermore, with the present state of computational resources, a purely numerical resolution of turbulent scales encountered in engineering problems is unfeasible. Consequently, almost all such investigations have to resort to some degree of modeling. Turbulence models are constitutive relations attempting to relate quantities of interest to flow parameters using assumptions and simplifications derived from physical intuition and observations. Reynolds Averaged Navier-Stokes based models represent the pragmatic recourse for complex engineering flows, with a vast majority of simulations resorting to this avenue. Despite their wide spread use, RANS-based models suffer from an inherent structural inability to replicate fundamental turbulence processes and specific flow phenomena. For the past few decades, there has been a large increment in the utilization of computational fluid dynamics (CFD) based simulations in the engineering design process. However, for novel designs, critical decisions or certification purposes, such numerical results are always supplement with experimental results. This is due to the lack of universal reliability of model predictions for different flows and disparate quantities of interest, leading to limited trust in RANS predictions. In this regard, the quantification of such uncertainties is of fundamental importance to enhance the predictive capabilities of RANS simulations as an engineering tool. This exercise establishes the degree of confidence in the model's predictions of different quantities of interest, for different flows under diverse conditions. 

The sources of the epistemic uncertainty inherent to RANS models are twofold: \textit{structural} and \textit{parameter} uncertainty. Structural uncertainty arises due to the inadequacy of the closure expression to represent the underlying physics in turbulent flows. In the recent past, Duraisamy and co-workers \citep{kdur} utilize a data-driven approach wherein full-field datasets are utilized to infer and calibrate the functional form of model discrepancies. This is augmented by machine learning algorithms to reconstruct model corrections. \citet{ling1} employ a data-driven approach along with a variety of machine learning algorithms to identify regions in the flow where high degrees of model-form uncertainty are extant. Similarly, turbulence model expressions use a number of closure coefficients which are usually assumed to be constant and are determined by calibration against a database of benchmark flows. The appropriation of the best-possible values of these coefficients, and, their functional form introduces additional uncertainty in the model framework. This is termed as parameter uncertainty. In this regard, \citet{edelinga} have investigated the parameter variability in the $k-\epsilon$ turbulence model. Similarly, \citet{margheri} and \citet{edelingb} have focused on investigating the sensitivity of the model predictions on the values of the model coefficients.

Recently, a physics-based, non-parametric approach to estimate the model-form uncertainties has been developed by \citet{emory1}. This framework approximates structural variability via sequential perturbations injected into the predicted Reynolds stress eigenvalues, eigenvectors and the turbulent kinetic energy. The eigenvalue perturbation formulation of this approach has been applied to engineering problems with considerable success \citep{gorle1,gorle2,gorle3,mishra2016}. In a similar vein, Xiao and co-workers \citep{xiao,xiao2,xiao3} have used a data-driven framework to inject uncertainty into the Reynolds stress eigenvalues to ascertain model form uncertainties.

However, a fundamental encumbrance in these studies arises due to the absence of a methodology to perturb the eigenvectors of the Reynolds stress tensor. For instance, the eddy-viscosity hypothesis common to mixing-length models, as well as one-, two- and three-equations closures like the Spalart-Allmaras, $k-\omega$ and $v^2-f$ models, assumes that the Reynolds stress has the same eigenvectors as the mean rate of strain. This is known to be invalid in complex flows, for instance, those involving streamline curvature, flow separation or rapid accelerations of the fluid \citep{speziale1990, launder1987, mishra2014}. The inconsistency in the eigenvector alignments between model predictions and experimental studies has been identified as a major source of discrepancy in the RANS modeling framework. For instance, the eddy-viscosity hypothesis common to mixing-length models, as well as one-, two- and three-equations closures like the Spalart-Allmaras, $k-\omega$ and $v^2-f$ models, assumes that the Reynolds stress has the same eigenvectors as the mean rate of strain. This is known to be invalid in complex flows, for instance, those involving streamline curvature, flow separation or rapid accelerations of the fluid.

In this investigation, we outline, validate and apply a comprehensive methodology to study the structural uncertainty in turbulence models. We focus on injecting uncertainty into the shape and the orientation of the Reynolds stress tensor to quantify the variability and biases introduced into the predictions. This framework attempts to assess the model uncertainty within the purview of single-point closures, without utilizing any additional data. While there have been excellent investigations, using data driven techniques to assess such perturbations for instance in \citet{xiao2}, we outline a purely physics-based approach. In this vein, the perturbations of the Reynolds stress anisotropy are governed by sampling from the extreme states of the turbulence componentiality. Similarly, the perturbations to the Reynolds stress eigenvectors are guided by the maximal states of the production mechanism. Jointly, these provide for a schema wherein the envelope of model-form uncertainty can be estimated within two simulations of the RANS model, minimizing the computational overhead. After introducing the mathematical foundations underlying this framework, as a proof of concept, we apply this methodology to a classical benchmark separated turbulent flow, over a backward-facing step, while contrasting the results against numerical and experimental data. Thence, this framework is applied to the high-speed flow in a nozzle to estimate the epistemic uncertainty in RANS predictions. As a datum for contrast, we explore the aleatoric uncertainty arising in the same configuration due to possible variation in the input parameters. 

\section{Mathematical rationale \& validation}
Turbulence models are constitutive relations attempting to relate unknown quantities (including higher-order statistical moments) to local, low-order flow quantities using simplifying assumptions and intuition, driven by analogies to physical processes. Such analogies are often of limited pertinence in complex flows, especially those encountered in industrial applications. In this section, we address two of the fundamental simplifications in this context, derived via an analogy of turbulent fluctuations to molecular motions: the \textit{Boussinesq hypothesis} and the \textit{gradient diffusion hypothesis}. The rationale underlying these simplifications, their limitations and ramifications are outlined. Thence, we demonstrate the manner in which sequential steps of the spectral perturbation approach seek to address the uncertainties introduced by these assumptions.

Most established RANS closures employ the Boussinesq hypothesis, relating the instantaneous values of the Reynolds stress tensor to the local rate of strain via:
\begin{equation}
R_{ij}\equiv\frac{2}{3}k\delta_{ij}-2\nu_{T}S_{ij},
\end{equation}
wherein the positive scalar coefficient, $\nu_{T}$, is the eddy-viscosity, computed using an assumed constitutive relation characteristic to the RANS closure. Herein the eddy viscosity is a scalar and thus, this constitutive relation between the fluctuating and mean fields is assumed to be the same in all directions. This aspect of the Boussinesq hypothesis may be a useful approximation in very simple shear flows, explicitly parallel shear flows where only one dominant shear direction is present. However, it is highly questionable in flows with strong separation, swirl or three-dimensional effects. This forced isotropy leads to poor approximations of the normal and shear stresses and therefore, a misrepresentation of the turbulence anisotropy structure. Additionally, the Boussinesq hypothesis obligates the modeled Reynolds stress to share its eigen-directions with the mean rate of strain tensor. This is patently erroneous for complex turbulent flows, especially those encountered in engineering applications. A detailed investigation into the limitations of the Boussinesq hypothesis in complex flows has been carried out by \citet{ling1}. The Boussinesq hypothesis is mathematically equivalent to the proportionality between the viscous stress and the rate of strain for Newtonian fluids. In fact, the original derivation of this relation is explicitly motivated by the kinetic theory of gases. This analogy is inappropriate due to significant differences in the motion of molecules in a gases and that of Lagrangian fluid particles in a turbulent flow. For instance, the turbulent mixing length is not small compared to the length scales of the mean flow variation, so that the requisite separation of scales does not hold for general turbulent flows as it does in the kinetic theory of gases. 

In a similar vein, the turbulent transport process is modeled using the gradient diffusion hypothesis, analogous to the Fick's law of molecular diffusion. For instance, considering the evolution equation for the turbulent kinetic energy:
\begin{equation}
\frac{\partial k}{\partial t}+U_i\frac{\partial k}{\partial x_i}=-\frac{\partial T_i}{\partial x_i}+\mathcal{P}-\epsilon,
\end{equation}
wherein $T=\frac{1}{2}\langle u_iu_ju_j \rangle+\frac{\langle u_i p \rangle}{\rho} -2\nu \langle u_j s_{ij} \rangle$, is the transport term for the turbulent kinetic energy. Using the gradient diffusion hypothesis, this is modeled as
\begin{equation}
T_i\equiv-\frac{\nu_{T}}{\sigma_{k}}\frac{\partial k}{\partial x_i}.
\end{equation}
Physically, this simplified model asserts that there is a flux of the turbulent kinetic energy along its spatial gradient, arising due to the turbulent fluctuations. A similar framework is applied in the modeled evolution equations for $\epsilon$, $\omega$, etc. The origins of this simplification lie in the mixing length model of Prandtl (1925). Herein, the fluid particles are envisioned to transport their momentum unchanged from their original location to their final location separated by a mixing length. Thence, with an analogy to the Fick's law of molecular diffusion (and the assumption that the turbulent mixing length, analogous to the diffusion length, is comparably small), this transport was modeled as gradient diffusion process.

The spectral perturbation approach introduced by \citet{emory1}, seeks to address these deficiencies in a systematic manner, to provide estimates for the uncertainties in the predictions of RANS closures. The perturbations are injected directly into the modeled Reynolds stress, ensuring the applicability of this estimation methodology to varied closures. The perturbed Reynolds stress can be expressed as:
\begin{equation}
R_{ij}^*=2k^* (\frac{\delta_{ij}}{3}+v^*_{in}\Lambda^*_{nl}v^*_{lj})
\end{equation}
wherein $R_{ij}^*$, represents the perturbed Reynolds stress; $k^*$, the perturbed turbulent kinetic energy; $v^*$, the perturbed eigenvector matrix, and, $\Lambda^*$, the diagonal matrix of perturbed eigenvalues. The tensors $v$ and $\Lambda$ are ordered such that $\lambda_1 \geq \lambda_2 \geq \lambda_3$. Each sequential step of this perturbation methodology seeks to address deficiencies in the RANS closure and the limitations due to the assumptions made therein. The eigenvalue perturbations address the limitation of the \textit{isotropic} eddy viscosity assumption. The eigenvalue perturbations treat the turbulent flow as an \textit{orthotropic} medium, assigning different turbulent viscosities along the three orthogonal axes defined by the eigen-decomposition of the rate of strain tensor. This enables the perturbed evolutions to exhibit better predictions of the secondary shear stresses and normal stresses. The eigenvector perturbations address the limitation of assuming that the eigen-directions of the instantaneous Reynolds stress tensor are always aligned with the local mean rate of strain. Mathematically, this accounts for additional degrees of anisotropy to the turbulent viscosity. Furthermore, if devised carefully, eigenvector perturbations can enable the inclusion of some aspects of non-local physics while determining uncertainty bounds. Finally, the perturbations to the turbulent kinetic energy explore the encumbrances due to the limitation in the representation of the turbulent transport process via the gradient diffusion hypothesis.

In this section, we outline a framework for the perturbation of the eigenvalues and the eigenvectors of the modeled Reynolds stress tensor. Adhering to a single-point formalism, this framework maximizes the information we can glean from single-point statistics, while minimizing the computational expense incurred in estimating uncertainty envelopes.

The perturbations to the eigenvalues, $\Lambda$, correspond to varying the componentiality of the flow, or equivalently, the shape of the Reynolds stress ellipsoid. These are defined through the coordinates in the barycentric map, $x$, via $\lambda^*_i=B^{-1}\mathbf{x}^*$, where perturbed quantities are starred and $B$ is the transformation from the eigenvalue space to the barycentric triangle. The projection of the eigenvalue perturbation in the barycentric map has both a direction and a magnitude. Considering the extreme states of Reynolds stress componentiality, we consider perturbation alignments to the three vertices of the triangle: $x_{1C}, x_{2C}, x_{3C}$ representing the $1C$, $2C$ and $3C$ limiting states of turbulence anisotropy. The magnitude of the perturbation in the barycentric triangle is represented by $\Delta_{B} \in [0,1]$. Thus, the perturbed barycentric coordinates, $\mathbf{x}^*$ , are given by: $\mathbf{x}^*=x+\Delta_B(\mathbf{x}^{(t)}-\mathbf{x})$, where $x(t)$ denotes the target vertex (representing one of the $1C$, $2C$ or $3C$ limiting states) and $x$ is the model prediction. Instead of relying on a user-defined perturbation for $\Delta_B$ , we set $\Delta_B=1.0$ so that the three limiting states are considered.

The perturbations to the eigenvectors, $v$, correspond to varying the alignment of the Reynolds stress ellipsoid. To guide these, we focus on the production mechanism, $\mathcal{P}=-R_{ij}\frac{\partial U_i}{\partial x_j}$. At the single-point level of description, this is the only turbulence process that is closed and does not require any simplifications or modeling. In our methodology, the eigenvector perturbations seek to modulate this production mechanism, by varying the Frobenius inner product $\langle A,R \rangle =tr(AR)$, where $A$ is the mean gradient and $R$ is the Reynolds stress tensor. For the purposes of bounding all permissible dynamics, we seek the extremal values of this inner product. In the coordinate system defined by the eigenvectors of the rate of strain tensor, the corresponding alignments of the Reynolds stress eigenvectors are given by $v_{max}=\begin{bmatrix}
  1 & 0 & 0 \\
  0 & 1 & 0 \\
  0 & 0 & 1
 \end{bmatrix}$  and $v_{min}=\begin{bmatrix}
  0 & 0 & 1 \\
  0 & 1 & 0 \\
  1 & 0 & 0
 \end{bmatrix}$. The corresponding ranges of the inner products are $[\lambda_1\gamma_3+\lambda_2\gamma_2+\lambda_3\gamma_1, \lambda_1\gamma_1+\lambda_2\gamma_2+\lambda_3\gamma_3]$, where $\gamma_1 \geq \gamma_2 \geq \gamma_3$ are the eigenvalues of the symmetric component of A.

\begin{figure}
\begin{center}
\includegraphics[width=1.05\textwidth]{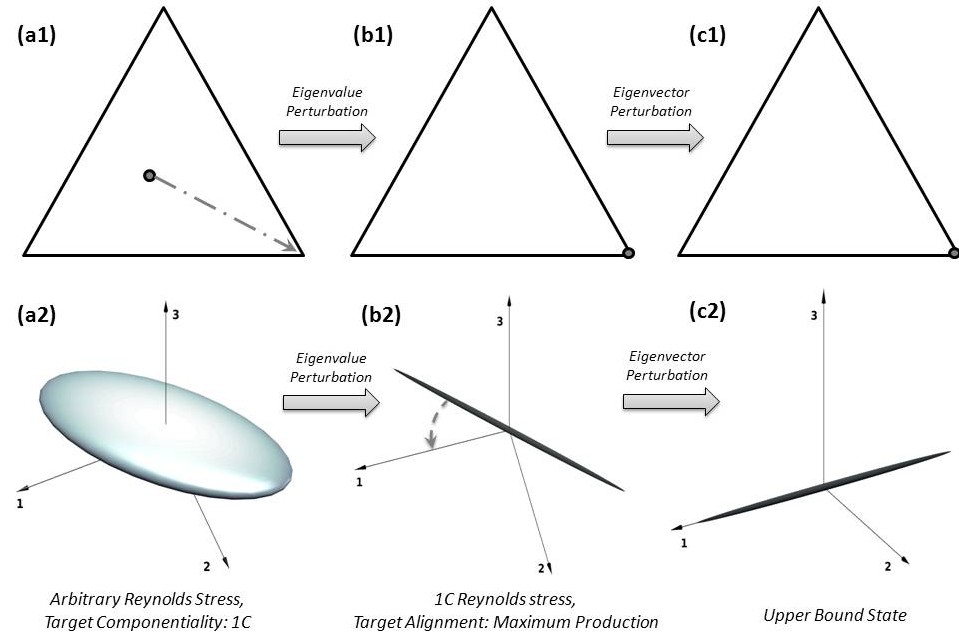}
\caption{Schematic outline of eigenvalue-eigenvector perturbations from an arbitrary state of the Reynolds stress. \label{fig:fig1}}
\end{center}
\end{figure}

To aid an intuitive feel for our perturbation methodology, we outline a representative case schematically in figure \ref{fig:fig1}. In the upper row, $(a_1,b_1,c_1)$, we represent the Reynolds stress tensor in barycentric coordinates and in the lower row, $(a_2,b_2,c_2)$, we visualize the Reynolds stress ellipsoid in a coordinate system defined by the mean rate of strain eigenvectors. These are arranged so that $\lambda_1 \geq \lambda_2 \geq \lambda_3,$ so essentially, the 1-axis is the stretching eigendirection and the 3-axis is the compressive eigendirection.
 
Initially, the Reynolds stress predicted by an arbitrary model is exhibited in the first column. The eigenvalue perturbation methodology seeks to sample from the extremal states of the possible Reynolds stress componentiality. Thus, we can, for instance, translate the Reynolds stress from a general $3$ component state to the $1C$ state, exhibited in the transition from $a_1$ to $b_1$. This translation changes the shape of the Reynolds stress ellipsoid from a a tri-axial ellipsoid to an extreme prolate ellipsoid, (in essence, a line) exhibited in the transition from $a_2$ to $b_2$.
 
Now, the eigenvector perturbation methodology, seeks to sample from the extremal states of the turbulence production mechanism, by varying the alignment of this ellipsoid. Thus, we can, for instance, rotate the Reynolds stress ellipsoid so that its semi-major axis is aligned with the stretching eigendirection of the mean rate of strain tensor, exhibited in the transition from $b_2$ to $c_2$. This particular alignment would enable us to analyze effects of the maximum possible production on flow evolution. 

Together, these two perturbation frameworks enable us to maximize the information we can get from single-point statistics to quantify uncertainty bounds, while implicitly accounting for certain features of non-local physics.
 \begin{figure}
\begin{center}
\includegraphics[width=1.05\textwidth]{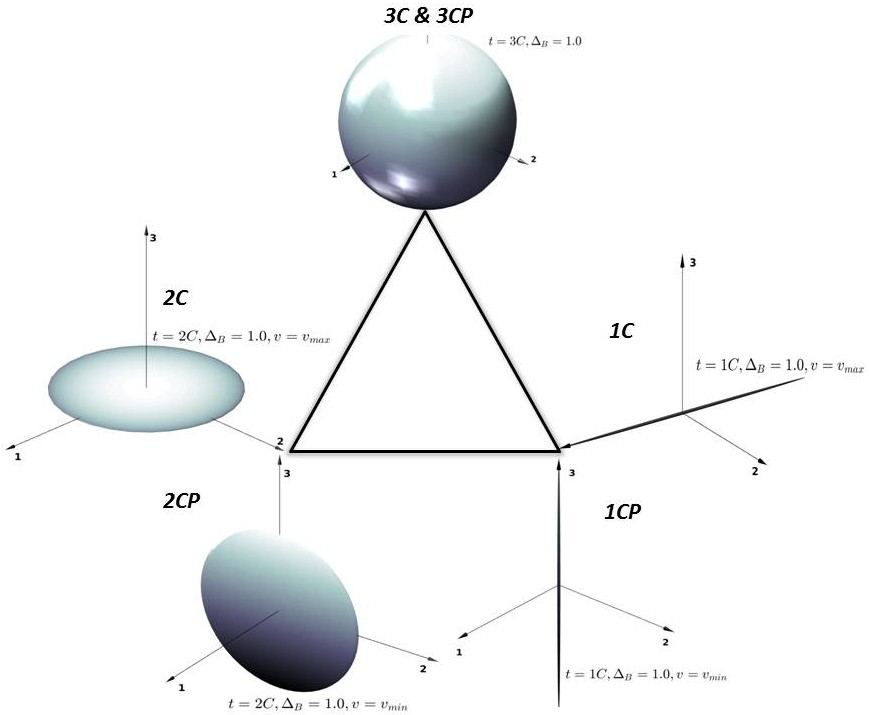}
\caption{Schematic visualization of the extremal states in the enveloping models methodology.\label{fig:fig2}}
\end{center}
\end{figure}

This eigenvalue-eigenvector perturbation framework gives us $5$ distinct extremal states of the Reynolds stress tensor, these are schematically displayed in figure \ref{fig:fig2}, along with their cognomen used in this investigation. These correspond to 3 extremal states of the componentiality $(1C, 2C, 3C)$ and 2 extremal alignments of the Reynolds stress eigenvectors, $(v_{min}, v_{max})$. For the $3C$ limiting state, the Reynolds stress ellipsoid is perfectly spherical. Due to its complete rotational symmetry at this state, all alignments of the spherical Reynolds stress ellipsoid are identical. Thus, for complete uncertainty bounds on flow evolution, we need a maximal set of only 5 RANS simulations. Further, we have found that for almost all cases, \textit{uncertainty bounds of engineering utility can be produced using just 2 specific perturbations}, labeled $3CP$ and $1C$, that seek to minimize and maximize production respectively. Producing uncertainty bounds from just 2 RANS simulations minimizes the computational cost associated with QMU. In summary, this eigenvalue-eigenvector framework \textit{maximizes the information we can glean from single-point statistics, while minimizing the computational expense}.

\begin{figure}
\begin{center}
\includegraphics[width=0.6\textwidth]{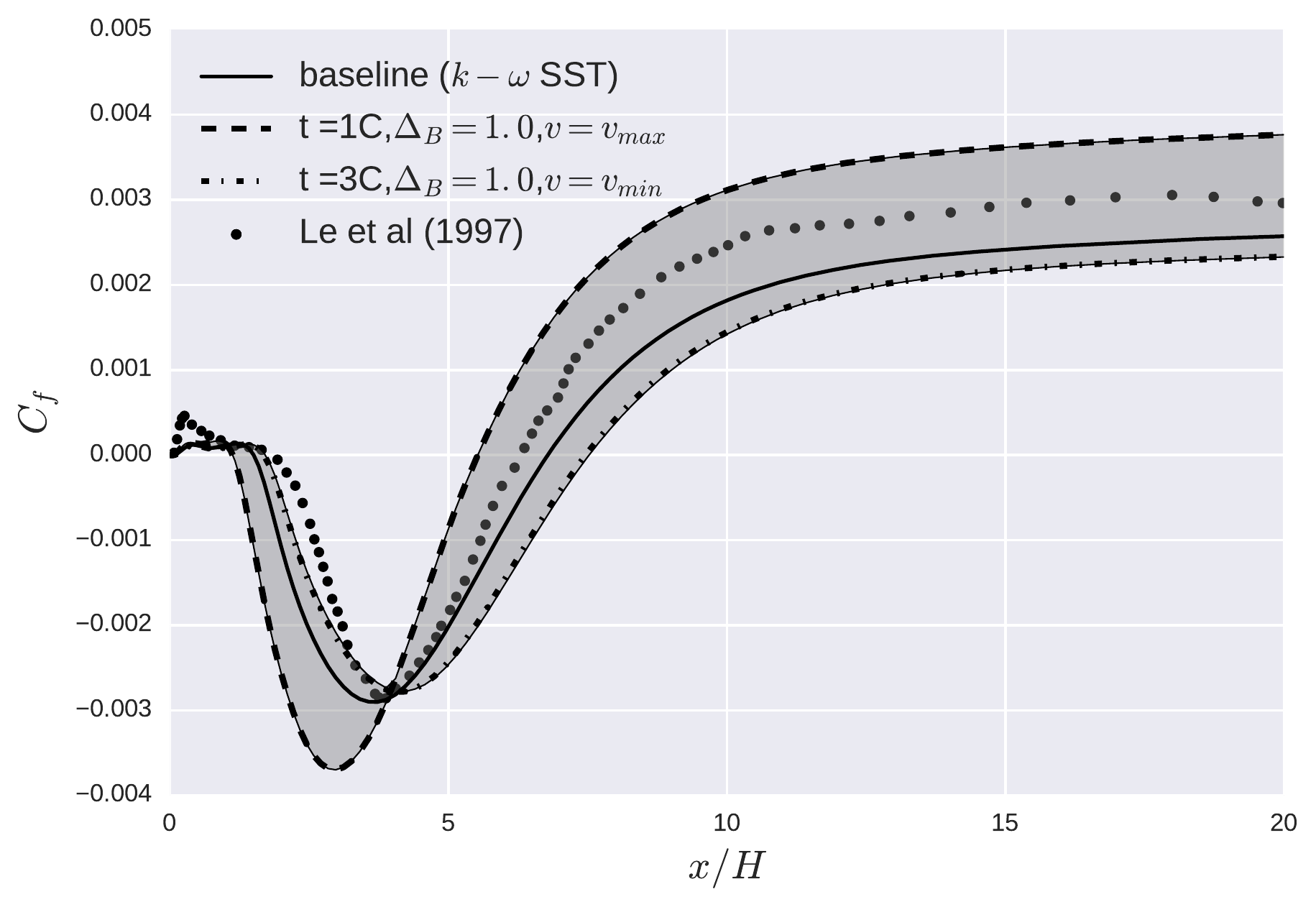}
\caption{Uncertainty envelopes on the skin friction coefficient, contrasted against DNS results of Le et al (1997)\label{fig:fig3}}
\end{center}
\end{figure}
\begin{figure}
\begin{center}
\includegraphics[width=0.7\textwidth]{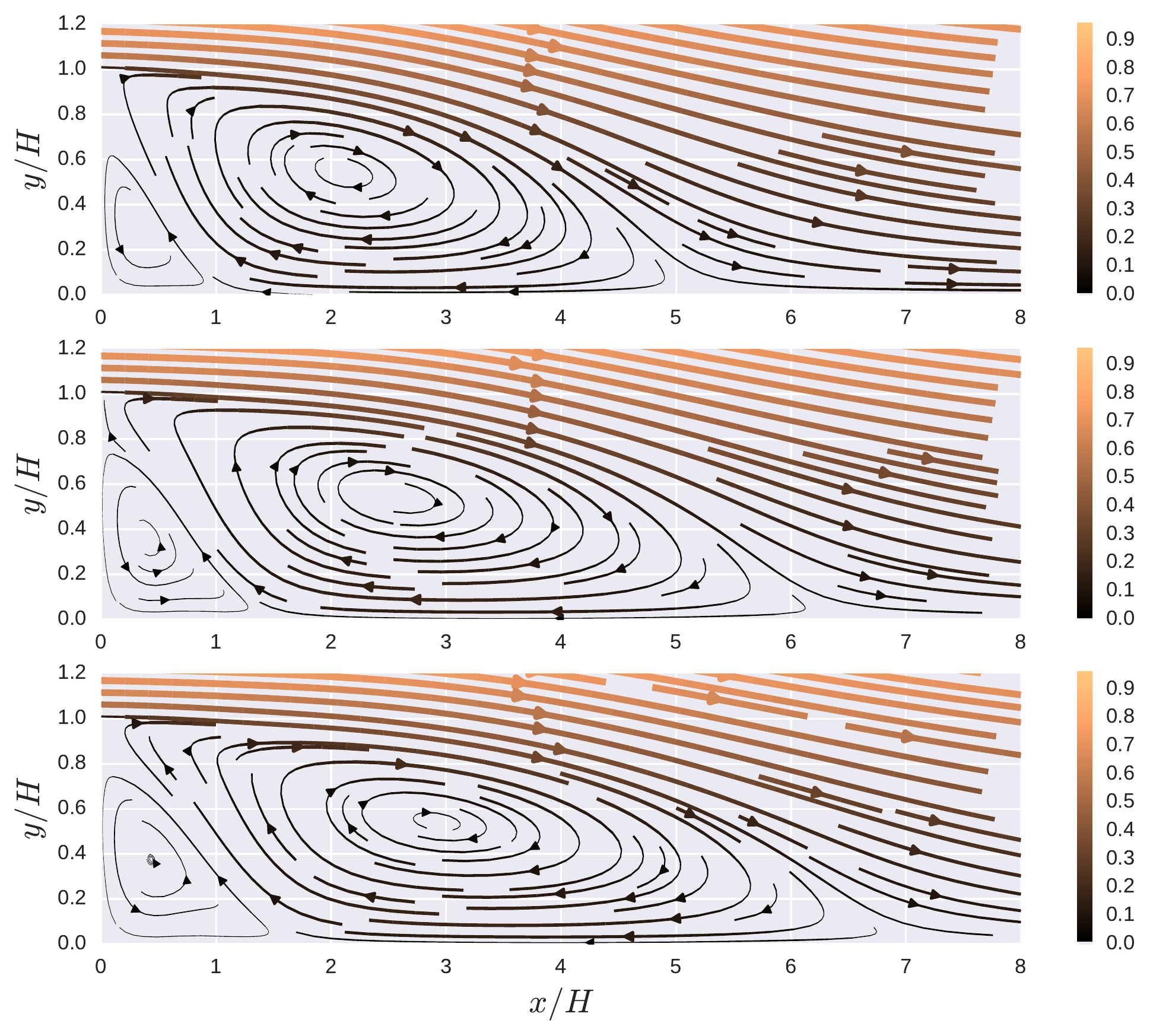}
\caption{Flow streamlines across the backward-facing step: (a) $(t = 1C, \Delta_B = 1.0, v = v_{max})$, (b) Baseline case, (c) $(t = 3C, \Delta_B = 1.0, v = v_{min})$.\label{fig:fig4}}
\end{center}
\end{figure}
For the validation of this methodology, we have applied it to a variety of separated turbulent flows, while contrasted against numerical and experimental data. The results for the canonical case of a turbulent flow over a backward-facing step are outlined herein, while utilizing the closure of \citet{sst}. In figure \ref{fig:fig3}, we consider the distribution of the skin-friction coefficient, $C_f$, along the bottom wall. The two RANS simulations corresponding to $(t = 1C, \Delta_B = 1.0, v = v_{max})$ and $(t = 3C, \Delta_B = 1.0, v = v_{min})$, capture the bounds on the possible behavior of the SST $k-\omega$ model. Figure \ref{fig:fig4} exhibits the flow streamlines in the zone of separation for this flow. As is suggested in the $C_f$ bounds in figure \ref{fig:fig3}, maximizing the production process suppresses the flow separation, while minimizing production leads to a broader zone of separated flow.

\begin{figure}
\begin{center}
\includegraphics[width=\textwidth]{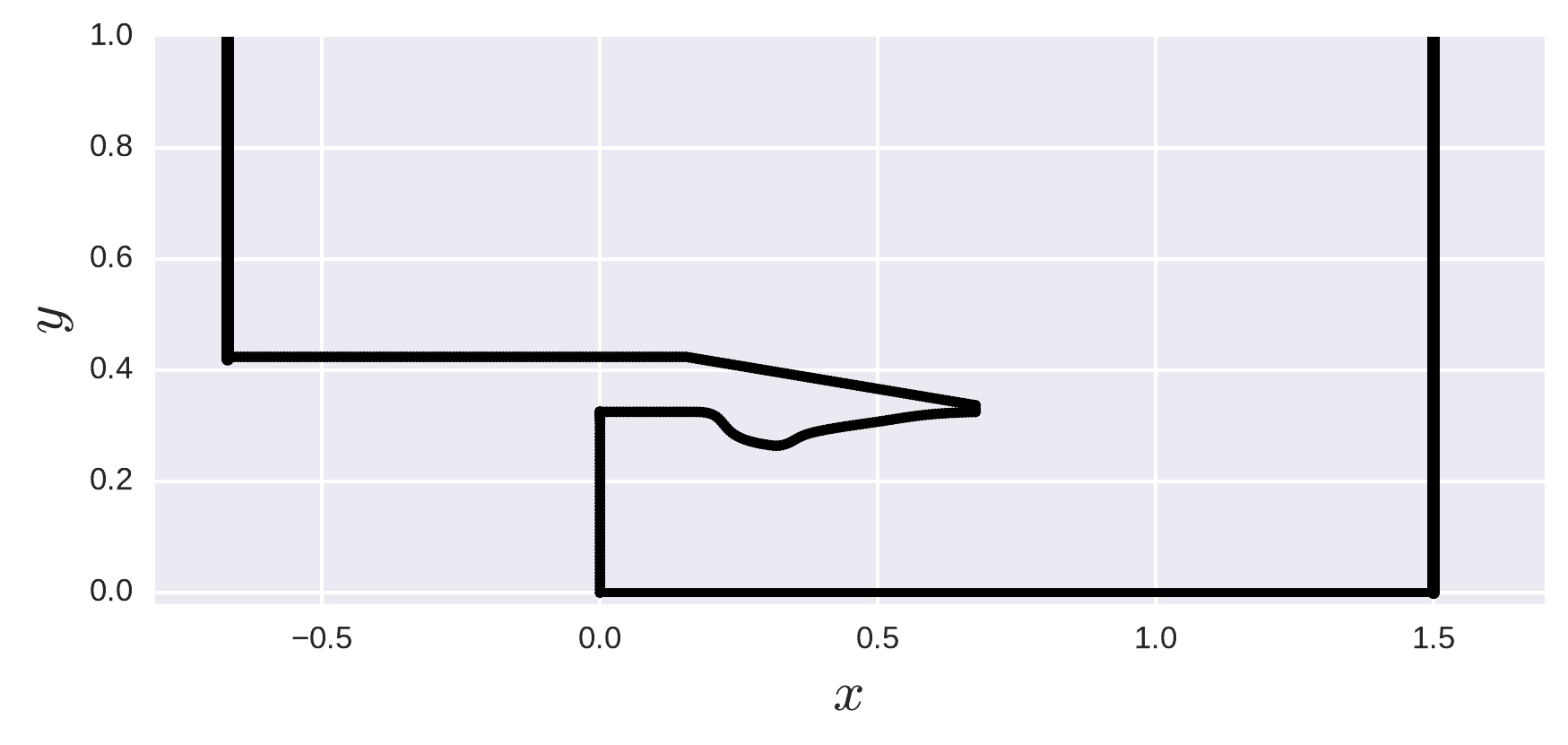}
\caption{Schematic detailing the hypersonic nozzle.\label{fig:fig5}}
\end{center}
\end{figure}

In the following section, the enveloping models methodology is used to ascertain the uncertainty bounds for a supersonic axisymmetric ``submerged" jet flow, while utilizing the SST $k-\omega$ RANS closure. The geometry is detailed in figure \ref{fig:fig5}. The computational geometry and flow conditions were adopted from the National Program for Application-oriented Research in CFD (NPARC) Alliance, a partnership between the NASA Glenn Research Center (GRC) and the Arnold Engineering Development Center (AEDC). The flow entails a Mach $2.22$ axisymmetric nozzle exhausting to the atmosphere and operating at the design pressure ratio, with the jet total temperature equal to ambient temperature. The internal circular-cross-section was constructed from a GRC design detailed in \citet{warren}, using the method of characteristics. This flow has been investigated via experiments in \citet{eggers}.

\begin{figure}
\begin{center}
\includegraphics[width=0.9\textwidth]{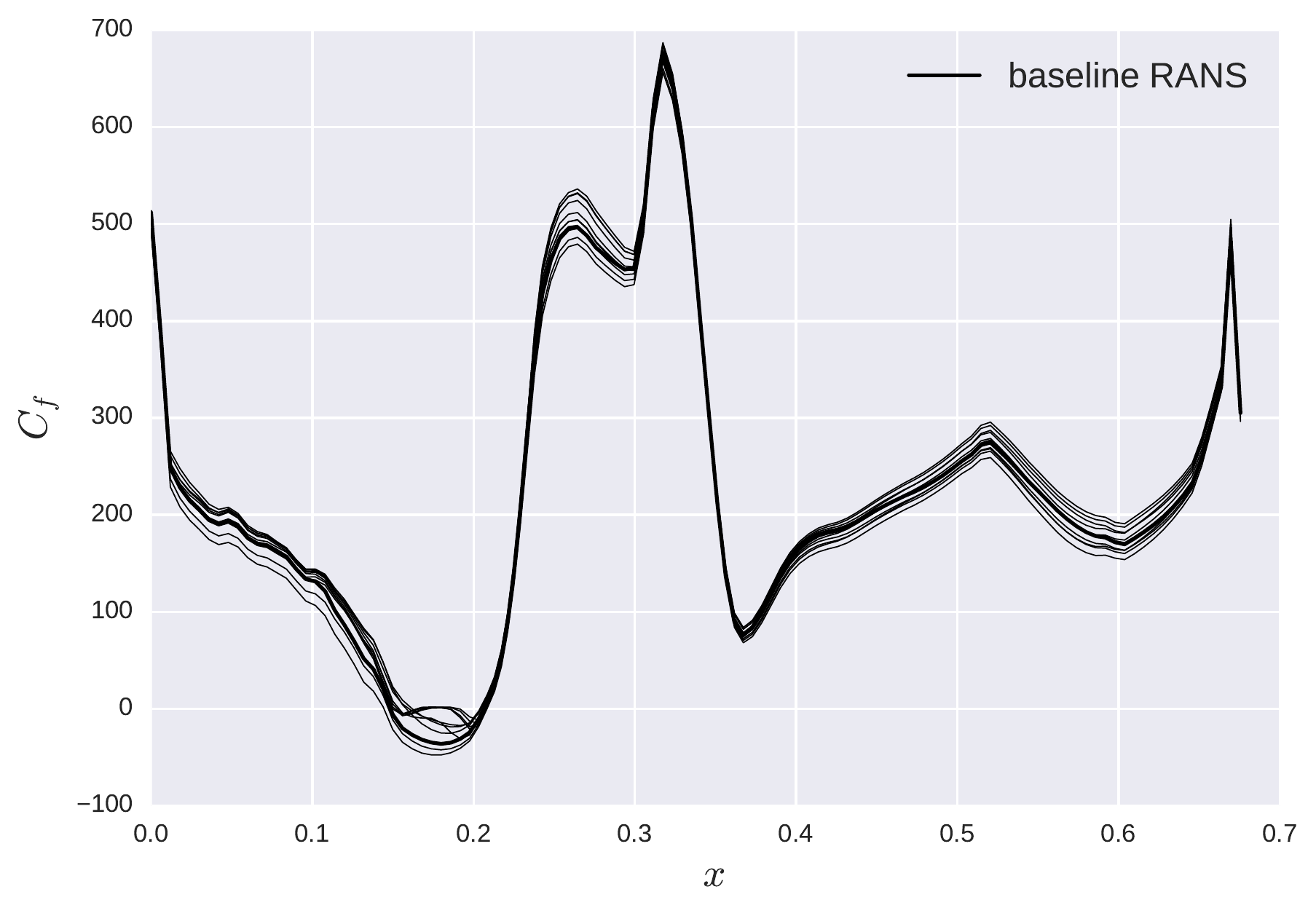}
\includegraphics[width=0.9\textwidth]{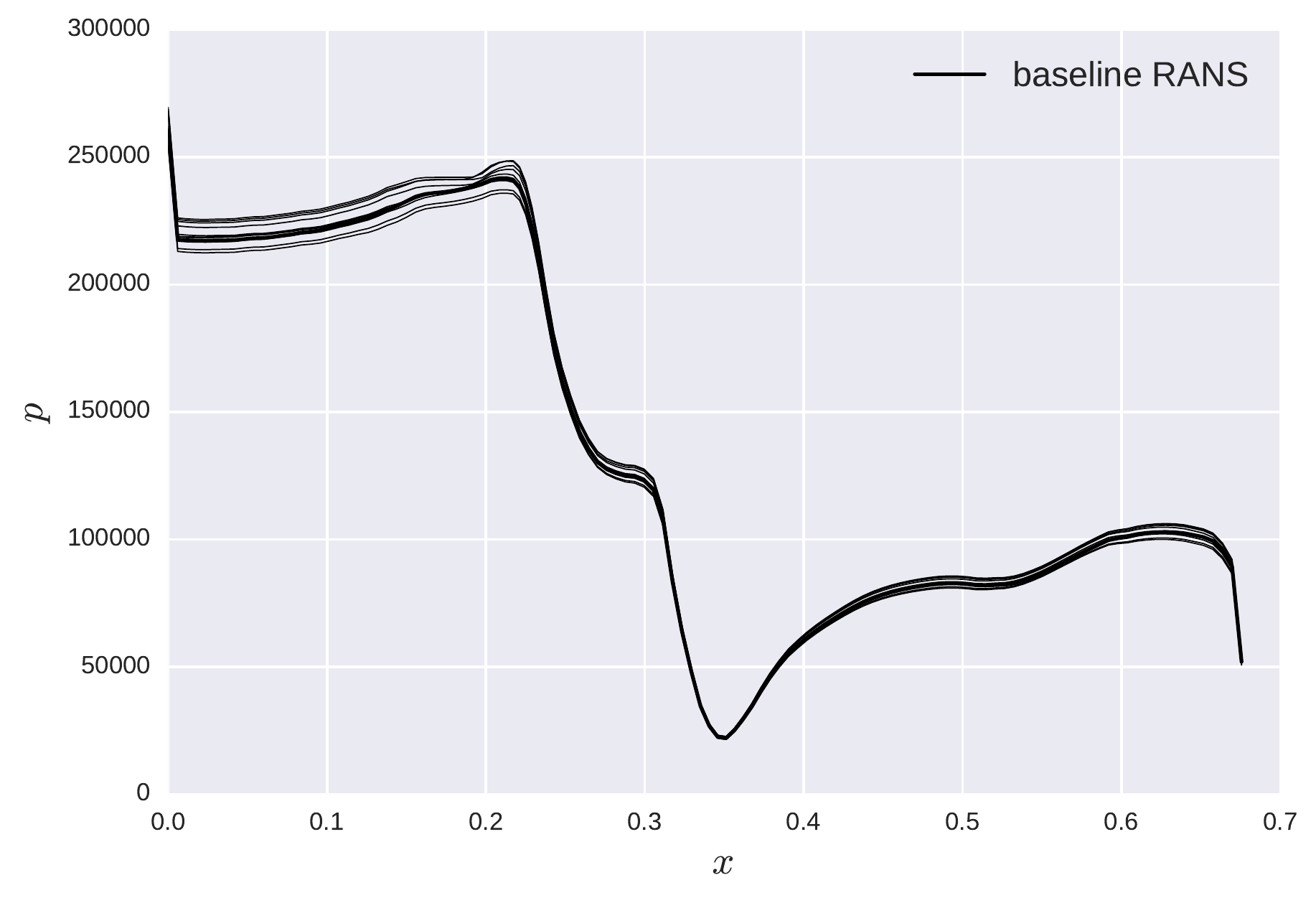}
\caption{Parametric uncertainty bounds on the predicted quantities of interest inside the nozzle: (a) Coefficient of friction, $C_f$, (b) Pressure, $p$. \label{fig:fig6}}
\end{center}
\end{figure}

\section{Application to high-speed flow}

\begin{figure}
\begin{center}
\includegraphics[width=0.7\textwidth]{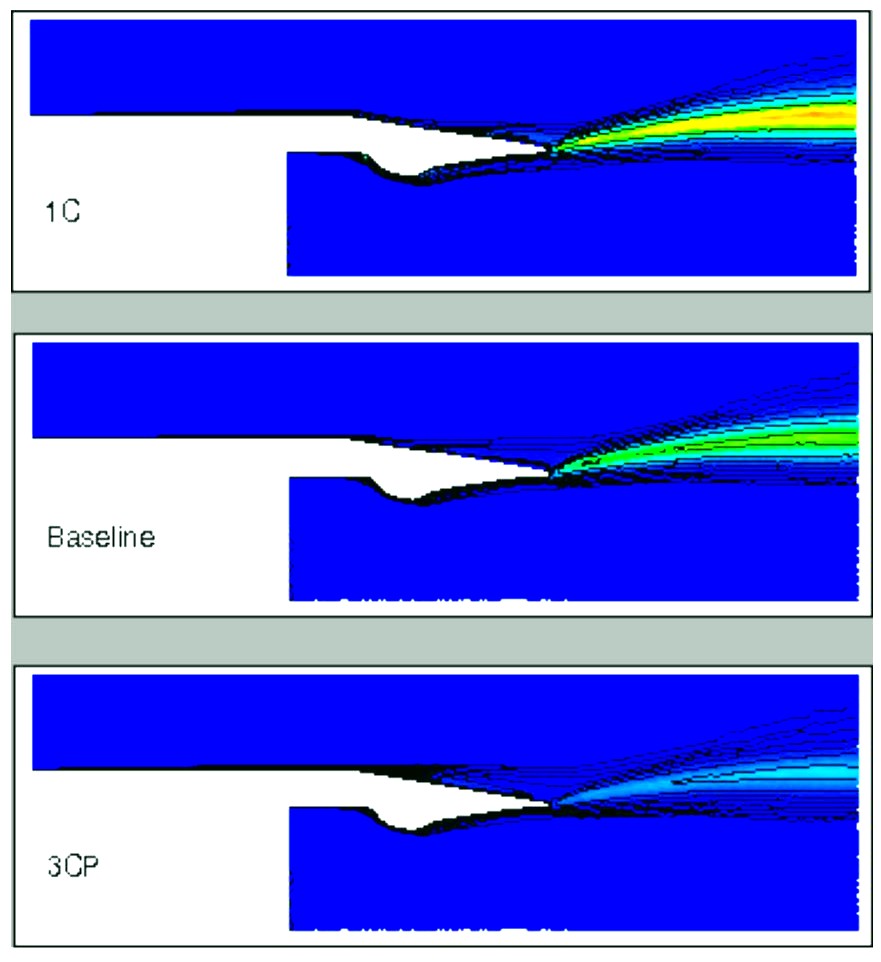}
\caption{Variation of turbulent kinetic energy, $k$, in the efflux region downstream of the nozzle outlet: (a) \textit{1C case} $(t = 1C, \Delta_B = 1.0, v = v_{max})$, (b) Baseline case, (c)\textit{ 3CP case} $(t = 3C, \Delta_B = 1.0, v = v_{min})$. \label{fig:fig7}}
\end{center}
\end{figure}

To ascertain a datum on the uncertainty on the quantities of interest in the flow within the nozzle, we commence with a series of simulations from perturbed initial conditions. Herein, the pressure and temperature values were varied over $\pm 5\%$ of their flow conditions. This series of Monte Carlo simulations provides for a measure of aleatoric uncertainty by quantifying the parametric variability on the prediction of the quantities of interest. The results for the Coefficient of friction, $C_f$, and Pressure, $p$ are detailed in the gitter plots in figure \ref{fig:fig6}. As can be seen, the RANS predictions are reasonably insensitive to small variations in the input variables to the model. The relatively narrow envelope of predictions from the perturbed simulations subsumes the baseline RANS prediction at all points inside the nozzle. Over the first section of the internal circular-cross-section, where the flow is diverging rapidly, there is a significant variation in the $C_f$ predictions, as should be expected. Similarly, over the latter half of the internal circular-cross-section, wherein the flow is contracting at a rapid rate, the effect of the variations is damped, as is expected again. 

Considering the epistemic uncertainty, in figure \ref{fig:fig7}, we outline the jet efflux from the nozzle in the free jet region. The jet from the nozzle interacting with the quiescent ambient fluid creates a velocity shear causing high degrees of turbulence and mixing. Accordingly, the contours in figure \ref{fig:fig7} correspond to the turbulent kinetic energy, $k$. As is expected from theory, the $1C$ case exhibits higher and broader distribution of $k$ in the efflux region. Similarly, the $3CP$ case has lower and narrower distributions of $k$. These two cases envelope the baseline SST $k-\omega$ predictions.

\begin{figure}
\begin{center}
\includegraphics[width=0.9\textwidth]{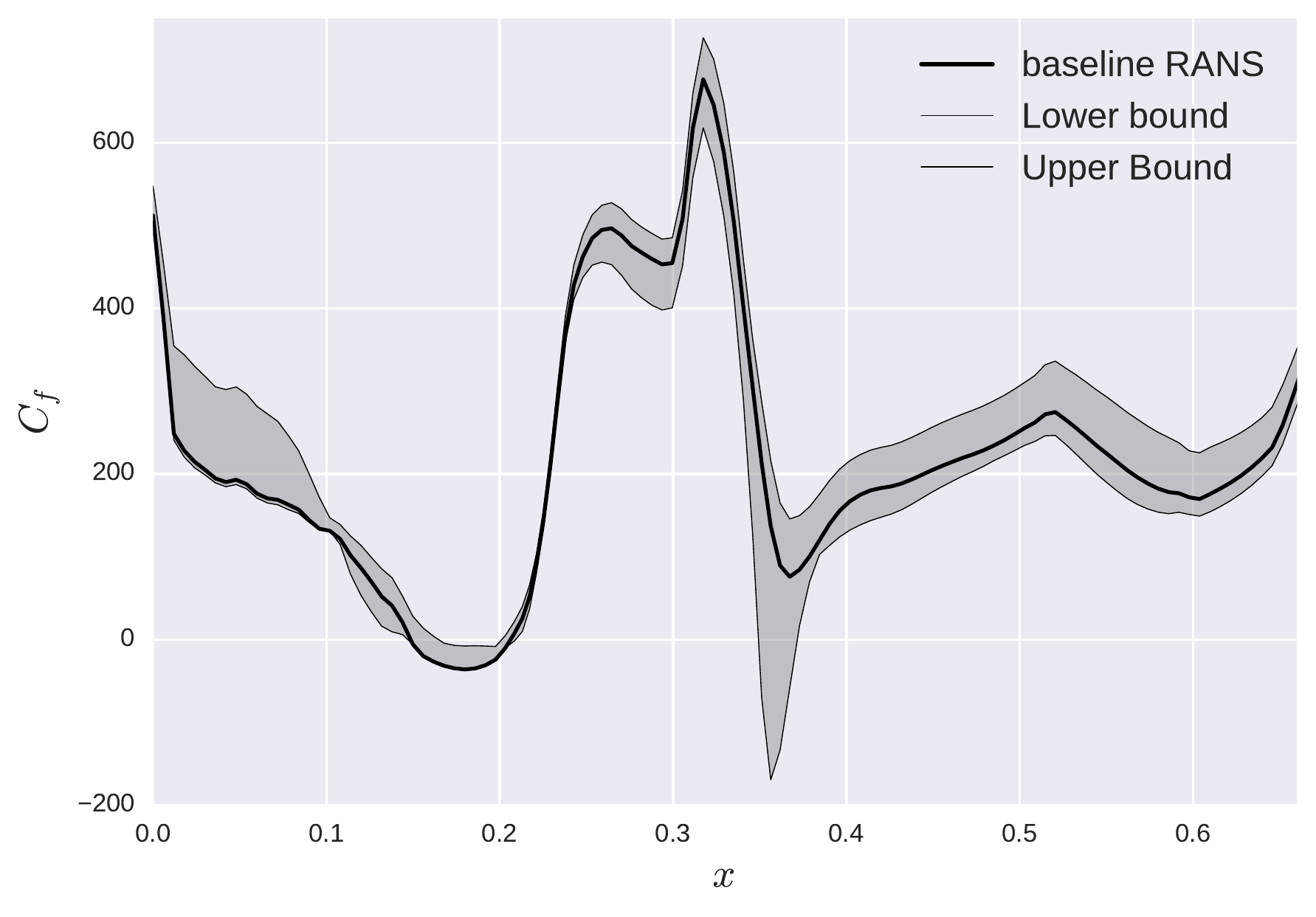}
\includegraphics[width=0.9\textwidth]{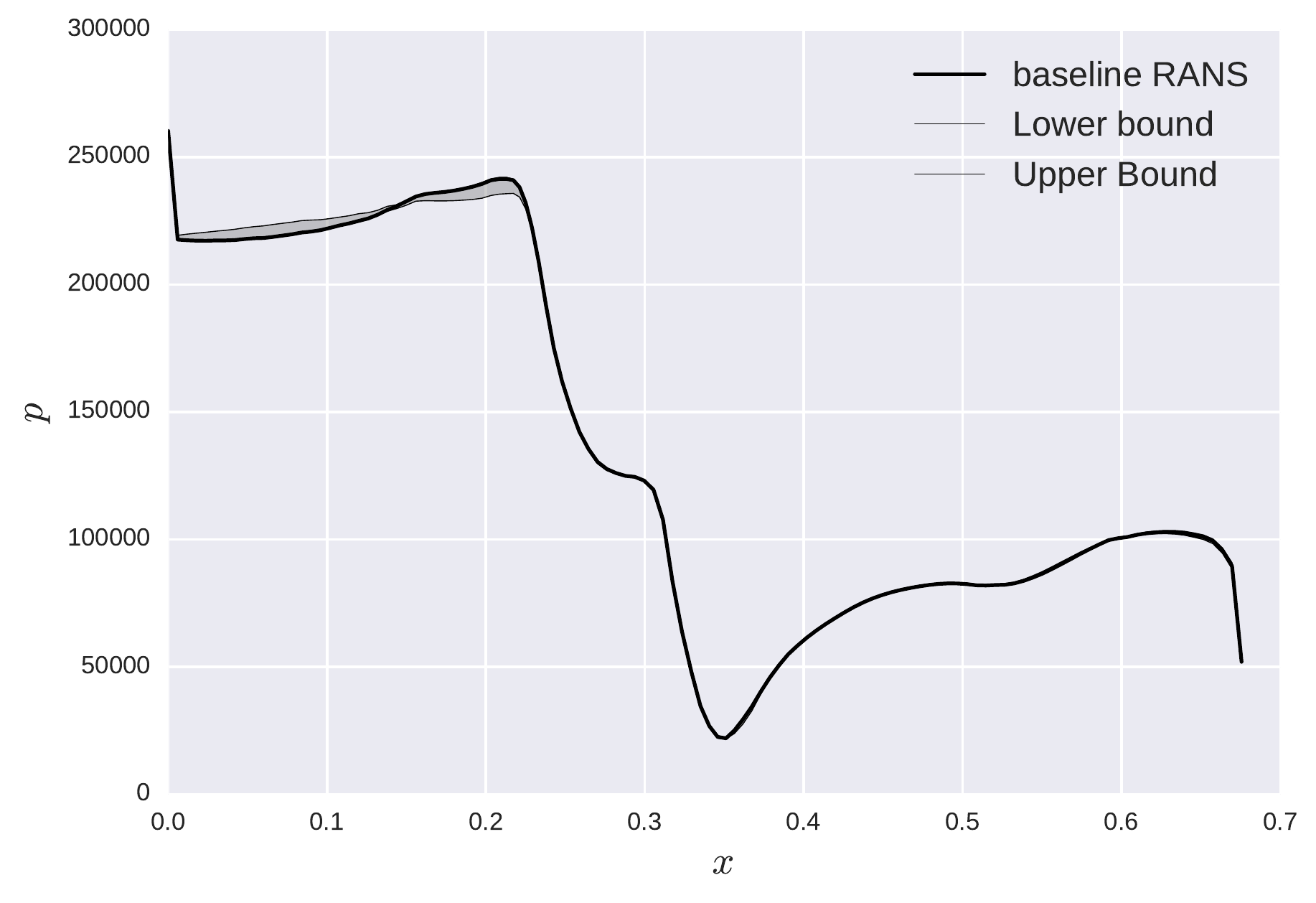}
\caption{Epistemic uncertainty bounds using enveloping models: (a) Coefficient of friction, $C_f$, (b) Pressure, $p$. \label{fig:fig8}}
\end{center}
\end{figure}

Figure \ref{fig:fig8}, exhibits the epistemic uncertainty estimates on the Coefficient of friction, $C_f$, and Pressure, $p$, inside the nozzle geometry. Contrasting against figure \ref{fig:fig6}, we can determine the difference in the nature of the aleatoric and the epistemic uncertainties in the flow through the nozzle. The epistemic uncertainty bounds on $C_f$ are much broader than their aleatoric counterparts. However, the enveloping models subsume the baseline model predictions at all points inside the nozzle. Additionally, the characteristic reduction and increment in the uncertainty bounds in the respective converging and diverging sections of the internal circular-cross-section are evident in this case as well. Contrarily, the epistemic uncertainty bounds on the predictions of $p$ are relatively negligible and are markedly lower than their aleatoric counterparts. The uncertainty in predictions of pressure do not seem to be amenable to variation in the production mechanism.  

\section{Conclusions}

In this investigation, we outline an \textit{enveloping models} methodology for estimating structural uncertainty bounds on RANS closures. This methodology incorporates both eigenvalue and eigenvector perturbations in the spectral representation of the Reynolds stress tensor. The underlying rationale of this enveloping models methodology are explicated in detail and the mathematical nuances are outlined. This methodology is validated via its application to a canonical case of separated turbulent flow, while contrasted against numerical data. It is exhibited that this procedure is able to provide prudent estimates
on the uncertainty of quantities of interest. Furthermore, uncertainty bounds of engineering utility can be engendered using just two specific RANS simulations, minimizing the computational overheads associated with uncertainty estimation. Thence, this methodology is applied to a supersonic axisymmetric ``submerged" jet flow. The epistemic uncertainty envelopes are contrasted against the aleatoric uncertainty bounds predicated upon the predictions of the RANS closure. The methodology is able to provide satisfactory results for this high-speed flow.
\FloatBarrier
\section*{Acknowledgment} 
This research was supported by the Defense Advanced Research Projects Agency under the Enabling Quantification of Uncertainty in Physical Systems (\textit{EQUiPS}) project (technical monitor: Dr Fariba Fahroo).

\bibliographystyle{ctr}
\bibliography{CTRReview2016}

\end{document}